\documentclass[twocolumn,showpacs,preprintnumbers,amsmath,
amssymb,superscriptaddress]{revtex4}
\usepackage{graphicx}

\begin{document}


\title{Singularities  of Hele-Shaw Flow and Shock Waves in Dispersive Media}

\author{Eldad Bettelheim}
\affiliation{James Frank Institute, Enrico Fermi Institute of the
University of Chicago,   5640 S. Ellis Ave. Chicago IL 60637}
\author{Oded Agam}
\affiliation{Department of Physics, University of Washington, Seattle,
WA, 98195} \author{Anton
Zabrodin \footnote{ Also at ITEP, B. Cheremushkinskaya 25, 117259
Moscow, Russia.}} \affiliation{Institute of Biochemical Physics,
Kosygina str. 4, 117334 Moscow, Russia}
\author{Paul Wiegmann\footnote{ Also at Landau Institute of
    Theoretical Physics, Moscow, Russia.}}
\affiliation{James Frank Institute, Enrico Fermi
Institute of the University of Chicago,   5640 S. Ellis Ave.
Chicago IL 60637}

\date{\today}

\begin{abstract}
We show that singularities developed in the Hele-Shaw problem have
a structure identical to shock waves  in dissipativeless
dispersive media. We propose an experimental set-up where
the cell is permeable to a  non-viscous fluid and study continuation
of the flow through singularities.  We show that a singular flow in
this, non-traditional cell is described by the Whitham equations
identical to Gurevich-Pitaevski solution for a regularization of shock
waves in Korteveg-de-Vriez equation.  This solution describes
regularization of singularities through creation of disconnected bubbles.
\end{abstract}


\maketitle

{\it 1. Introduction.} A broad class of non-equilibrium growth
processes in two dimensions are characterized by a common law: the
velocity of the growing interface is determined by the gradient of
a harmonic field  (often referred to as Laplacian growth)
\cite{review}.


Growth processes  determined by a harmonic field, where no cut-off
scale is introduced, are important. The theory of this kind of
growth is deeply related to fundamental aspects of conformal maps,
integrable systems, 2D quantum gravity and random matrices
\cite{ourpapers}. However, singularities makes this problem
ill-defined and thus not achieved experimentally.


The goal of this Letter is two-fold. One is to suggest an
experimentally achievable set-up, where a high-rate flow can
continue through singularities, without being curbed by
dissipative forces. Another is to stress a deep and important
parallel between the singularities of growing interface and
``gradient catastrophes'' known in the theory of shock waves in
dispersive media. The Korteveg-de-Vriez (KdV) and other nonlinear
waves equations feature this phenomenon.  This relation allows
one to effectively study complicated singular processes.


The relation  between the Hele-Shaw problem and dispersive
nonlinear waves suggests  a unique  regularization,
namely one which plays a similar role as dispersion of
nonlinear waves. There, dispersion, (no matter how small it is)
becomes  crucial near a singularity, converting shock waves to
oscillatory structures \cite{singular}.

{\it 2. The Hele-Shaw flow.} The Hele-Shaw cell is a narrow space
between two plates filled by incompressible viscous liquid (say
oil).  Air (regarded as inviscid and incompressible)
occupies a part of the cell forming one or several bubbles.
Traditionally air is pumped into one bubble, while oil is
extracted from the cell at a constant rate $Q>0$ through the edges
placed at infinity. Without surface tension, the interface may
develop singular cusps at a finite time \cite{review}. At this
moment the problem becomes ill-posed.

\begin{figure}
\includegraphics[width=5.cm]{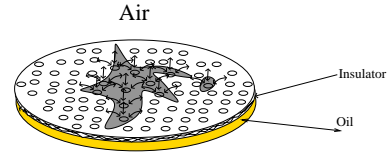}
\caption{\label{hele} Schematic Hele-Shaw experimental set up.
Shaded region represents a small bubble of air in the ambient oil.
The insulator is permeable to air but not to oil.} \end{figure}

We suggest a novel set up where the upper plate  is permeable to
air  and is connected to a reservoir of air. Neither plates are
permeable to oil. This set-up may be achieved by sticking a
Gortex-like material to glass plates having small perforations
(Fig.\ref{hele}). We thus assume that all air bubbles are kept at
the same atmospheric pressure.   We also consider  a retraction
problem,  where oil is extracted from the cell,  i.e. $Q<0$.

In both cells the flow obeys the same local equations (D'Arcy's
law): The local velocity of oil, averaged over the gap between the
plates, is proportional to a gradient of pressure $p$. In an
incompressible fluid pressure is a harmonic function.
Thus the normal velocity  of the interface, $v_n$, is  proportional
 to the normal derivative of pressure, and in proper units
$v_n =- \, \partial _n p$. If in
addition surface tension is ignored, the pressure is continuous
across the interface, and may be set to $p=0$ within each bubble. Thus
the pressure  solves  the exterior Dirichlet boundary value problem:
$\Delta p =0$, with $p=0$ on the boundary, and $p \to - Q\log |z|$
at infinity.

\begin{figure}
\includegraphics[width=7.cm]{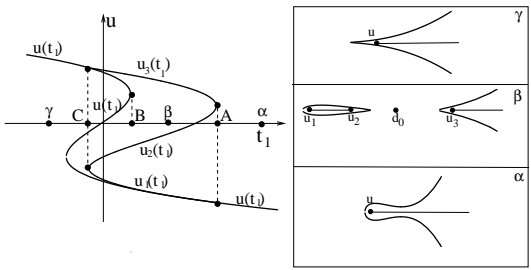}
\caption{\label{quality} Graph of numerical solution for branch
points $u_i(t_1)$ compared to the hodograph for  $u(t_1)$. Finger
shapes at early ($\gamma$), singular ($\beta$) and late ($\alpha$)
stages of evolution.} \end{figure}

{\it 3. Bubble break-off and merging.}
Let us now describe an air retraction process  in our experimental
set up.
A typical local evolution of  
an air bubble consists of few phases  illustrated in
Fig.~\ref{quality}: (i) As oil is injected, the air bubble
contracts; (ii) The air bubble forms a singular narrow neck and
breaks-up into one or several disconnected
bubbles; 
(iii) All bubbles subsequentially contract, while loosing air
through the upper plate.


As we show in this Letter, the method of regularization of shock
waves provides a detailed description of retraction evolution in
the limit of zero surface tension. This is in contrast with the
injection process in a traditional Hele-Shaw cell where solutions
at the zero surface tension limit are valid only until a cusp is
formed.  This difference, suggests that surface tension is a
non-singular perturbation in our new experimental setup.



\begin{figure}
\includegraphics[height=3.5cm]{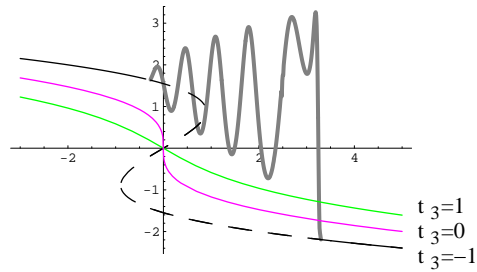}
\caption{\label{hodograph} Solutions of the  hodograph for
$t_3=-1$ (green), $0$ (red) and $1$ (black). Unphysical
multivalued region (in dashed line)  of the $t_3=-1$-graph is
replaced by an oscillatory solution in heavy gray line
(numerical).} \end{figure}

{\it 4. Dispersive regularization of shock waves.}
The term shock wave is usually used to describe nonlinear waves
in presence of dissipation.
In the absence of dissipation the shock-wave behavior
is different. It is resolved  by converting to oscillations. The
method to study this complicated behavior was suggested by
Gurevich and Pitaevskii (GP) \cite{GP}, and refined in later works
\cite{DN,F,singular}. They  \cite{GP} studied solutions to the KdV
equation
\begin{equation}\label{K} \partial_{t_3}u =\frac{3}{2}
u\, \partial_{t_1}u +\varepsilon \partial_{t_1}^3 u,
\end{equation}
with step-like boundary conditions: $u(t_1\to-\infty)=u_0$ and
$u(t_1\to+\infty)=0$. Initially the function $u(t_3,t_1)$ is
smooth ($t_3=1$ of Fig.\ref{hodograph}), so the dispersion term
$\partial_{t_1}^3 u$ may be neglected. However, the Hopf-Burgers
equation $\partial_{t_3}u =\frac{3}{2} u\partial_{t_1}u$, thus
obtained, always develops a shock wave, a singularity with  an
infinite slope ($t_3=0$ of Fig. \ref{hodograph}) followed by an
unphysical overhang ($t_3=-1$ of Fig. \ref{hodograph}). This
signals that the limit $\varepsilon\to 0$ is singular.

In fact the solution with finite $\varepsilon$ never develops a
shock wave. Before the singularity occurs the wave breaks into
fast oscillations (see Fig. \ref{hodograph}), of a period scaling
with $\varepsilon$.

When  $\varepsilon\ll 1$, the oscillatory regime can be described
by slowly modulated periodic solutions of the KdV equation. These
are given by the elliptic function \begin{equation}\label{2}
u(t_3,t_1)\approx2 \alpha\cdot \rm{dn}^2 \left( \frac{5
\sqrt{\alpha} }{ 12\sqrt{6} \varepsilon}(t_1 + V
t_3),m\right)+\gamma, \end{equation} which moduli  $\alpha,\,
m,\,V$,  $\gamma$ (i.e. frequency, amplitude etc.)  depend on the
times  $t_3,\,t_1$.  This dependence is described by the Whitham
equations \cite{Witham}. These equations appear below in (\ref{U},\ref{G16})
to describe  the Hele-Shaw flow.



A similar situation takes place in our problem.
We will show
that  a typical flow is identical to averaging the solution
(\ref{2}) over fast oscillations, and is, essentially, described
by the same Whitham equations.

{\it 5. Correspondence between interface dynamics and shock-wave
solutions.} In order to establish a link between modulated periodic solutions
of the KdV equation and growth of planar domains let us recall
the notion of the spectral curve \cite{book}.  The spectral curve
(or Riemann surface)  encodes periodic solutions of  nonlinear
integrable equations. In the KdV case it is a hyperelliptic curve
\begin{equation}\label{G5} y^2 = R_{m} (z),\quad m=2\ell+1,
\end{equation}
where $y$ and $z$ are complex variables, and  $R_{m}$ is a
polynomial of an odd degree,  such that  its roots are real. The
spectral curve of solution (\ref{2}) is given by a polynomial of
degree $m=5$ and  eq. (\ref{G14}). A non-periodic solution at
small $\varepsilon$ is described by a slowly varying spectral
curves.

A shock wave indicates a singular evolution when the curve changes
its genus. We will show that for the interface dynamics
an increase of the genus implies a bubble break-off, since
the interface is a real section of the curve when the coordinates
$z$ and $y$ are real\cite{KM-WWZ}-\cite{footnote}.


{\it 6. The critical flow.}
Once injecting air into Hele-Show cell, the  bubble develops a
``finger''. Its tip  is pushed away with increasing velocity that
eventually may result in a cusp-like singularity
(Fig.~\ref{quality}).

Let us choose the origin  at the point where a cusp would form,
and simplify the argument assuming that the finger is symmetric
with respect to its $x$-axis. In dimensionless units, where the
size of the entire droplet is of order 1 we have $|y|\ll |x| \ll
1$.  Let us denote the distance between the tip and the origin by
$u(t)\ll 1$. It sets the only (time dependent) scale of  the
critical flow.

A critical flow of an isolated finger is conveniently formulated
as the time evolution of the Cauchy transform of air bubbles:
\begin{equation}\label{G3}
h^{(\pm) }(z)=\frac{1}{2\pi i}\oint _{\gamma} \frac{y' dz'}{z -
z'},\quad z=x+iy.
\end{equation}
This integral defines analytic functions $h^{\pm }(z)$ for $z$
inside and outside air respectively. A direct calculation shows
that D'Arcy law is equivalent to \begin{equation}\label{G4}
\partial _t h^{(+)}(z)=0, \quad \partial _t h^{(-)}(z)=-2i
\partial _z p(z, \bar z) \end{equation} The first equality implies
an infinite series of conserved quantities first observed in
\cite{R}.

A generic  form of the critical finger (insets $\gamma$ and
$\alpha$ in Fig.~\ref{quality}) is given by a  curve (\ref{G5}),
$y^2=R_m(x)$, such that  all the coefficients of the polynomial
$R_m(x)$ are real and it has at least one real root $x=u$ at the
tip of the finger.   In the critical regime the coefficients of
the polynomial $u^mR_{m}(x/u)$ are of order 1, while $u\to 0$ as
time approaches the critical time $t_c$. There a finger becomes
the $(2,2\ell+1)$ cusp $y^2\sim |x|^{2\ell+1}$.

The critical flow is conveniently  described in terms of the
height function \cite{TWZ}.  The height function
$y(z)=\sqrt{R_m(z)}$  is defined on a hyperelliptic Riemann
surface, and is an analytic function outside the finger, having
boundary value $y(x)$ on the boundary of the finger. In \cite{TWZ}
it was proved that: (i) the finger remains self-similar, i.e.,
remains of the form of polynomial of a fixed degree, only if all
its  simple roots are real. In this case   branch points other
than $u$ correspond to additional finite size droplets, if any. As
a consequence, a sole finger is described by a degenerate curve
$\sqrt{R_m(z)}=\sqrt{z-u}\,P_{\ell}(z)$, where $P_{\ell}$ is a
polynomial of degree $\ell$; (ii) The coefficients $t_3,  \ldots ,
t_{2\ell +3}$ (called deformation parameters) in front of all
positive fractional powers of $z$  in the expansion of
\begin{equation}\label{G10} y(z)=\sqrt{R_m (z)}=\sum _{k=0}^{\ell
+1} (k +\frac{1}{2}) \, t_{2k+1}z^{k-\frac{1}{2}} + O(z^{-3/2})
\end{equation} do not depend on time.
Furthermore, the coefficient in front of the first negative power,
$z^{-1/2}$, is proportional to time $t_1 \sim  Q (t-t_c )$. The
negative tail of the series (\ref{G10}) depends on time in a
non-trivial way.

For simplicity let us consider a single finger and study
the behavior of $h^{\pm}(z)$ in the domain $|u|\ll |z|\ll 1$,
far from the tip but around the finger, where details
 of the tip are not seen.
The Cauchy integral (\ref{G3}) for $z>u$ on the positive real axis
contains only regular terms (i.e. positive integer powers of $z$).
Therefore, $h^{(+)}$ extends analytically as a regular function to
the whole domain of interest. The  singular part of the function
$h^{(-)}$ (containing also fractional powers of $z$) then provides
all necessary information. It has a cut inside the finger drawn
along the $x$-axis (see Fig.\ref{quality}).

In order to compute the Cauchy integral (\ref{G3}), we expand the
function $y(x)$ in a series in half-integer powers of $x$ and
evaluate $h^{(-)}(z)$  term by term. The function $h^{(-)}(z)$ has
a cut along the finger axis from $u$ to infinity and takes
opposite real values on the two sides of the cut. Conventionally,
we choose a single-valued branch such that $y(x+i0)>0$. For real
negative $z=x$ we find $h^{(-)}(x)= y(x)$. Therefore, the singular
part of $h^{(-)}(z)$ is to be identified with the height function
\begin{equation}\label{G8} h^{(-)}(z)= y(z)=\sqrt{R_m (z)}
\end{equation} Now consider the evolution of the finger (or the
curve (\ref{G5})) encoded by Eq.(\ref{G4}). The leading behavior
of pressure around the finger follows from solution to the
Dirichlet boundary value problem around a slit  (since $|y|\ll
|x|$, the finger roughly looks like a cut): $p(z, \bar z) \propto
{\rm Re} \, \sqrt{-z}$. This proves that higher terms in the
expansion of the height function  (\ref{G10}) are conserved.

{\it 7. Dispersionless KdV hierarchy.}  We have reformulated the
Hele-Shaw flow as an evolution of a hyperelliptic curve. Notably,
the  spectral curve of KdV  equation (\ref{K}) evolves  exactly in
the same way. In particular,  it follows from (\ref{G4}-\ref{G8})
that  before the break-off  the scale $u$ evolves with time and
other deformation parameters according to the dispersionless KdV
hierarchy \cite{TWZ} $$ {2^{n}n!}\partial_{ t_{2n+1}} u+
{(2n+1)!!}u^n \partial_{ t_1}u=0.
$$

{\it 8. (2,5) critical finger.}  A generic flow is represented by
a family of critical fingers of the form (\ref{G5}) with
$2\ell+1=5$ \begin{equation}\label{G12} y^2
=(x-u)(x-d_{+})^2(x-d_{-})^2.
\end{equation}
The $x^4$-term, and therefore $t_5$ in (\ref{G10}), can always be
eliminated by a shift ($2d_{+}+2d_{-}+u =0$), leaving $t_3$ as
the only deformation parameter. Substituting this to (\ref{G10})
we obtain the hodograph solution \cite{hodograph,hodographnote}
\begin{equation}\label{hod} \frac{5}{8}\, u^3 + \frac{3}{2}\, u
t_3 + t_1 =0, \end{equation} implicitly giving the branch point
$u$ in terms of  $t_1$, $t_3$.\\ \indent A singularity occurs when
$u$ merges with one or two double points
$d_{\pm}=\frac{1}{4}(-u\pm \sqrt{-24t_3 -5u^2})$.

The features of finger's evolution crucially depend on the sign of
$t_3$ (Fig. 3). If $t_3>0$, then the double points never reach the
real axis.  The function $u(t_1)$ is single-valued and  in both
extraction and injection processes the  finger never becomes
singular (inset $\alpha$ in Fig.\ref{quality}).

In the case $t_3=0$ the solution is $u(t)\propto - (Q(t-t_c
))^{1/3}$. This corresponds to a finger which evolves into the
$(2,5)$-cusp $y^2 = x^5$. At $t=t_c$, all the three roots
coincide: $u=d_{+}=d_{-}=0$. An interesting feature of the (2,5)
cusp (and of all cusps $(2, 2\ell +1)$ at even $\ell$) first noted
in \cite{Howison} is that the evolution can be extended beyond the
cusp by means of the same hodograph equation (\ref{hod}).

The most interesting case is $t_3<0$. A formal solution in the
injection case leads to the (2,3)-cusp, $y^2\propto x^3$, which
can not be continued. The plot in Fig. 3 becomes multi-valued in
the region $t_{1}^2 < \frac{4}{5}(-t_{3})^3$.

{\it 9. Air extraction -- bubble break-off.} We treat air
extraction, where $Q<0$, and the physical time is $- t_1$.
At an early stage, the finger tip is far to the left ($u$ is large
negative and $t_1$ is large positive). Extracting air from the
bubble results in a tip motion from left to right and a changing
shape of the finger in accordance with the hodograph solution. The
evolution follows the branch of the plot from $t_1 =-\infty$ until
the point $A$ in Fig. \ref{quality}. At this stage, the double
points $d_\pm$ of the curve (\ref{G12}) are complex. At $t_1 =
t_A\equiv \sqrt{{54}/{5}} \, (-t_3 )^{3/2}$ they approach the real
axis and merge. As $t_1 \to t_A$, the finger develops a thin neck
around the point $-u/4$, which breaks at $t_1 = t_A$ through the
(2,4) cusp $y^2\propto x^4$. At the break-off the curve is $y^2
=(x-u)(x+\frac{1}{4}u)^4$ ($u$ is negative).  The bubble which
breaks off from the main one has area
$\frac{25\sqrt{5}}{84}(-u)^{7/2}$. After this singular point, the
evolution according to the hodograph solution (\ref{hod}) is
unphysical. It leads to an interface with a self-intersection
point.

The actual evolution at $t_1< t_A$ does not follow the cubic
parabola (\ref{hod}) in Fig.3. The correct extension of the
solution beyond the singularity describes a small bubble breaking
off from the finger tip. A physical solution describes the
interface (or a spectral curve) $y^2 = R_5 (x)$ with $R_5 (x)$
having only one rather than two double roots:
\begin{equation}\label{G14}
y^2 =(x-u_1)(x-u_2) (x-u_3)(x-d_{0})^2,
\end{equation}
where $d_0=V=\frac{1}{2}(u_1+u_2+u_3)$. Now the height function
$y(z)$ has two cuts: one small cut inside the small bubble and an
infinite cut inside the finger (inset $\beta$ in Fig.2). The
double point $d_0$ moves between them. Mathematically, this means
that the complex curve is of genus 1 (an elliptic curve).
Similarly to the genus-0 case (\ref{G12}),  a substitution
(\ref{G14}) into (\ref{G10})   we obtain
\begin{equation}\label{U}
12 t_1 = U_{1}^{3}-4U_3\,, \quad
 -12 t_3 = U_{1}^{2}+2U_2,
\end{equation}
where $U_k =u_{1}^{k}+u_{2}^{k}+u_{3}^{k}$.

In case of two bubbles, $t_1$, $t_3$ are not enough to fix the
dynamics. An additional conserved quantity follows from the
condition on pressure. Since air is under the same pressure in
both bubbles, we  write $\int_{u_2}^{u_3}dp =2{\rm Re}\,
\int_{u_2}^{u_3}\partial _z p \, dz = 0$, or, using (\ref{G4}),
${\rm Im}\, \int_{u_2}^{u_3}\partial _t h^{(-)}dz =0$. Since
$h^{(-)}(u_i)=y(u_i)=0$, we  rewrite this condition as $\partial
_t \left ( {\rm Im}\, \int_{u_2}^{u_3} \sqrt{R_5 (x)}\, \, dx
\right ) =0$. The quantity under the integral is purely imaginary
and vanishes at $t_1= t_A$, hence
\begin{equation}\label{G16}
\int_{u_2}^{u_3} \sqrt{R_5 (x)}\, \, dx =0.
\end{equation}
This condition  closes the system of equations (\ref{U}) after the
break-off. However, unlike the algebraic equations for $t_k$, this
condition is transcendental. A solution is available through
elliptic functions. It gives the time dependence of functions
$u_1, \,u_2,\,u_3$ plotted in Fig.\ref{quality}. After the
break-off, the small bubble starts to evaporate and eventually
disappears at $t_1= t_C= - \, \sqrt{{2}/{27}} \, (-t_3 )^{3/2}$
corresponding to the point $C$ on the plot. After that the
solution switches back  to the cubic parabola. The finger proceeds
further without obstacles.

{\it 10. Discussion.} Below we highlight our major results. We
propose a novel,  experimentally accessible modification of the
Hele-Shaw cell, where the flow proceeds through singularities
without being curbed by  a surface tension. In this cell
singularities of the flow are resolved by creation of new bubbles,
all kept at the same constant pressure. This behavior can be seen
experimentally.

The flow in such an air-permeable cell can be studied with the
help of mathematical tools of soliton theory. We identified
bubble break-off with the shock wave behavior of the KdV
equation.

We are grateful to A.Marshakov,
S.Nagel, R.Te\-odo\-res\-cu and H.Swinney for discussions. We  are
particularly grateful to I.Krichever  for help and contributions.
P.W. and E.B. were supported by the NSF MRSEC Program under
DMR-0213745 and NSF DMR-0220198. A.Z. was also supported in  part
by  grants RFBR 03-02-17373, INTAS 03-51-6346, NSh-1999.2003.2.
O.A. acknowledges support from the Israel Science Foundation (ISF)
and from the German Israel Foundation (GIF).


\begin{thebibliography}{99}
\bibitem{footnote}
The  concept of  an evolving Riemann surface can be traced back to
Richardson \cite{R}, who related the Hele-Shaw problem to
quadrature domains. The latter were then related to algebraic
Riemann surfaces \cite{Quadratures1,Quadratures2}, while a direct
application of this theory  was made in \cite{Quadratures3}
\end{thebibliography}
\end{document}